\documentclass{pasj00}
\draft
\begin{document}
\SetRunningHead{Tatsuya. Kamezaki et al}{VLBI Astrometry of RX~Boo.}
\Received{2010/07/12}
\Accepted{2011/08/11}

\title{VLBI Astrometry of the Semiregular Variable RX~Bootis}

\author{
Tatsuya~\textsc{Kamezaki}\altaffilmark{1}, 
Akiharu~\textsc{Nakagawa}\altaffilmark{1}, 
Toshihiro~\textsc{Omodaka}\altaffilmark{1}, 
Tomoharu~\textsc{Kurayama}\altaffilmark{1}, 
Hiroshi~\textsc{Imai}\altaffilmark{1}, 
Daniel~\textsc{Tafoya}\altaffilmark{1}, 
Makoto~\textsc{Matsui}\altaffilmark{1}, 
Yoshiro~\textsc{Nishida}\altaffilmark{1}, 
Takumi~\textsc{Nagayama}\altaffilmark{2}, 
Mareki~\textsc{Honma}\altaffilmark{2}, 
Hideyuki~\textsc{Kobayashi}\altaffilmark{2}, 
Takeshi~\textsc{Miyaji}\altaffilmark{2}, 
Mine~\textsc{Takeuti}\altaffilmark{3}
}
\altaffiltext{1}
{Department of Physics and Astronomy, Graduate School of Science and Engineering, Kagoshima University, 1-21-35 Korimoto, Kagoshima 890-0065, Japan}
\altaffiltext{2}
{Mizusawa VLBI Observatory, Mitaka Office, National Astronomical Observatory of Japan, 2-21-1 Osawa, Mitaka, Tokyo 181-8588, Japan}
\altaffiltext{3}
{Astronomical Institute, Tohoku University, 6-3 Aramaki, Aoba-ku Sendai, Japan 980-8578}

\email{kamezaki@milkyway.sci.kagoshima-u.ac.jp}
\maketitle

\begin{abstract}
We present distance measurement of the semiregular variable RX~Bootis (RX~Boo) with its annual parallax. 
Using the unique dual-beam system of the VLBI Exploration of Radio Astrometry (VERA) telescope, we conducted astrometric observations of a water maser spot accompanying RX~Boo referred to the quasar J1419+2706 separated by $1^{\circ}$.69 from RX~Boo.
We have measured the annual parallax of RX~Boo to be $7.31 \pm 0.50$\,mas, corresponding to a distance of $136^{+10}_{-9}$\,pc, from the one-year monitoring observation data of one maser spot at $V_{{\rm LSR}}=3.2\,{\rm km\,s}^{-1}$.
The distance itself is consistent with the one obtained with Hipparcos.
The distance uncertainty is reduced by a factor of two, allowing us to determine the stellar properties more accurately.
Using our distance, we discuss the location of RX Boo in various sequences of Period-Luminosity (PL) relations. 
We found RX~Boo is located in the Mira sequence of PL relation.
In addition, we calculated the radius of photosphere and the mass limitation of RX~Boo and discussed its evolutionary status.
\end{abstract}
\section{Introduction}
\label{sec:section1}
In some kinds of variable stars, there is relations between the absolute magnitudes and the variation periods.
This is called the Period-Luminosity (PL) relation.
PL relations of long period variable stars (for example, Miras or Semi-Regular Variable stars (SRVs)) have long been investigated since they were found in Large Magellanic Cloud (LMC) \citep{1981Natur.291..303G, 1989MNRAS.241..375F, 1996MNRAS.281.1347G, 2000PASA...17...18W, 2001A&A...377..945C, 2004MNRAS.353..705I, 2004MNRAS.348.1120N}.
When we use PL relation, we can estimate absolute magnitudes using periods.
With these absolute magnitudes and apparent magnitudes from observations, we can calculate distances of these stars.
Detailed study of PL relations has given the multiplicity of the relation usually explained by the difference in excited pulsation mode. 
For the red long-period variable stars, such a multiplicity is first reported by \citet{2000PASA...17...18W}.

Although PL relations in our galaxy are studied \citep{2000MNRAS.319..759W, 2002MNRAS.334..498Z, 2003A&A...403..993K, 2004MNRAS.355..601Y, 2007MNRAS.378.1543G, 2008MNRAS.386..313W}, more studies will be required to establish the precise relation.
In the case of LMC, we can study PL relations by using apparent magnitudes on assumption that  all variable stars in LMC have the same distances, since the thickness of LMC is much smaller than the distance to LMC.
On the other hand, it is obvious that variable stars in our galaxy have large relative difference in distances.
To determine more precise PL relations in our galaxy, we need to obtain distances and apparent magnitudes for more variable stars.
So, we measured the distance from annual parallax using a VLBI method.

Among long-period variable stars, RX Bootis (RX Boo) is one of the most interesting star because the star has two different periods \citep{1987JBAA...97..277T, 1988AN....309..323A, 1997ESASP.402..269M, 2006JAVSO..35...88S} and classified as SRb group which show low amplitude and less regular variability than Miras or SRa group. 
It is known that regular and high amplitude variables are found on a common PL relation for our galaxy and LMC (e.g. \citet{2008MNRAS.386..313W}). 
It is interesting whether or not the positions of RX Boo situate on the PL sequence of Miras.
The study of the parallax of RX Boo will be useful to clarify the nature of SRb group. 

We performed astrometric observations with the VLBI Exploration of Radio Astrometry (VERA).
VERA is a Japanese VLBI array dedicated to phase referencing VLBI observations \citep{2003ASPC..306..367K}.
It consists of four antennas (Mizusawa, Iriki, Ogasawara and Ishigakijima).
VERA can measure annual parallaxes and proper motions of many astronomical objects at 22 and 43\,GHz. 
With the dual-beam system of VERA, we observe target and reference sources simultaneously.
The system makes it possible to cancel out the effect of atmospheric fluctuations between two sources \citep{2008PASJ...60..935H}. 

\citet{2008A&A...482..831W} monitored RX~Boo over twenty years with single-dish observations using Effelsberg 100-m and Medicina 32-m telescopes. 
They also observed water masers in RX~Boo using the Very Large Array (VLA) on four occasions in the period from 1990 to 1992 and later in 1995.
Then they revealed the distribution of water masers around RX~Boo. 
They detected the emission of incomplete shell around RX~Boo. 
The inner radius of the shell is estimated to be 15\,AU.
They conclude that the variability of water masers around  RX~Boo is due to the appearance and disappearance of maser clouds with a lifetime of $\sim$1\,year. 
With infrared interferometers, the angular diameter of RX~Boo is estimated as $18.4\,\pm\,0.5$ mas and $21.0\,\pm\,0.3$ mas in $K$ band (2.2 \,$\mu$m) and $L^{\prime}$ band ($3.8\,\mu$m), respectively, with the model of uniform disks \citep{1996AJ....111.1705D, 2002AJ....124.2821C}.
We can convert these apparent sizes to actual sizes by using distance.
RX~Boo emits SiO and OH masers as well as water masers \citep{2004AAS...205.1207B, 1995A&A...297..494S}.
The annual parallax of RX~Boo was measured to be $6.42\pm1.00$ mas by Hipparcos \citep{1997A&A...323L..49P}.
From CO observations, \citet{2002A&A...391.1053O} estimated its mass-loss rate and expansion velocity of $6\,\times\,10^{-7}\MO\,{\rm yr^{-1}}$ and $9.3\,{\rm km\,s}^{-1}$, respectively.
\citet{2006A&A...450..167T}  also obtained the values  of $2\,\times\,10^{-7}\,\MO\,{\rm yr^{-1}}$and $7.5\,{\rm km\,s^{-1}}$ from CO observations.

In this paper, the observations and data reduction are described in section \ref{sec:section2}.
In section \ref{sec:section3}, we present the annual parallax of RX~Boo. 
Finally, in section \ref{sec:section4}, we discuss the PL relations and stellar properties of RX~Boo. 
 
\section{Observations and data reduction}
\label{sec:section2}
\subsection{VLBI observations}
\label{sec:section2.1}

\begin{table}[t]
\caption{Observation dates\label{tb:table1}}
\begin{tabular}{lcccc}
\hline
epoch&Date&Year/DOY&antenna$^{\dagger}$&$\Delta V$(${\rm km\,s}^{-1}$)$^{\ddagger}$\\
\hline
$1$& 2008 February 19&2008/037&4&0.42\\
$2^{*}$& 2008 May 1&2008/121&4&0.21\\
$3^{*}$& 2008 June 11&2008/162 &4&0.21\\
$4^{*}$& 2008 July 16&2008/197 &4&0.21\\
$5^{*}$& 2008 November 11&2008/315 &4&0.21\\
$6^{*}$& 2008 December 8&2008/342 &4&0.21\\
$7^{*}$& 2009 January 10&2009/010 &3&0.42\\
$8^{*}$& 2009 February 4&2009/035 &4&0.21\\
$9^{*}$& 2009 March 12&2009/071 &3&0.42\\
$10^{*}$& 2009 May 5&2009/125 &4&0.42\\
$11$& 2009 September 5&2009/248 &4&0.42\\
$12$& 2009 October 3&2009/276 &4&0.42\\
\hline
\multicolumn{5}{l}{* Observations used in the estimation of annual parallax.}\\
\multicolumn{5}{l}{$\dagger$ Total number of antennas joined the VLBI observation.}\\
\multicolumn{5}{l}{$\ddagger$ Velocity resolution of the maser IF for each observation.}
\end{tabular}
\end{table}

We conducted monthly VLBI observations of water masers from February 2008 to October 2009 with VERA.
We show the observation dates in Table \ref{tb:table1}.
The duration of each observation was typically 8\,hours, yielding an on-source integration time of 5 to 6 hours.
Synthesized beam size (FWHM) was typically 1.2\,mas$\times$0.6\,mas.
To obtain positions of maser spots around RX~Boo, we observed a continuum source J1419+2706 simultaneously as a position reference.
The phase tracking center of RX~Boo and J1419+2706 are 
$(\alpha_{J2000.0}$,\,$\delta_{J2000.0})$=(\timeform{14h24m11s.6206},+\timeform{25D42'12''.909}) and 
$(\alpha_{J2000.0}$,\,$\delta_{J2000.0})$=(\timeform{14h19m59s.2971},+\timeform{27D06'25''.5530}), 
respectively. 
The separation angle between two sources is $\timeform{1D.69}$. 
In three observations, we did not detect any maser spot due to the bad weather conditions or the time variation of maser emission.
Among twelve epochs, we used nine epochs that we detected a maser spot for the estimation of the annual parallax. 
In the seventh observation in 2009 January, Iriki station did not participate.
In the nineth observation in 2009 March, Ogasawara station did not participate.
The shape of the synthesized beam of nineth observation is different from those of the other observations.
Data recording rate of 1024\,Mbps was adopted with the VERA DIR2000 recording system, which yields a total bandwidth of 256 MHz with 2-bit digitization.
The 256\,MHz data of left-hand circular polarization were divided into 16 IFs which had bandwidth of 16\,MHz.
One IF is used to receive the maser emission and the others are used to receive the continuum emission from J1419+2706. 
The correlation was carried out with Mitaka FX correlator \citep{1998ASPC..144..413S} at National Astronomical Observatory of Japan (NAOJ).
The rest frequency of 22.235080\,GHz is adopted for water maser emission. 
The velocity resolution ($\Delta V$) of the maser IF for each observation is also shown in table \ref{tb:table1}. 
In six observations, the IF assigned to the water maser was divided into 512 spectral channels, yielding a frequency resolution of 31.25\,kHz, corresponding to a velocity resolution of 0.42\,km\,s$^{-1}$.
In the other observations, the maser IFs have bandwidths of 8MHz.
They were divided into 512 spectral channels, yielding a frequency resolution of 15.625\,kHz and a velocity resolution of 0.21\,km\,s$^{-1}$.
For the data of J1419+2706, each IF channel was divided into 64 spectral channels in all observations. 
This corresponds to the velocity resolution of  3.37${\rm~km\,s}^{-1}$.

\subsection{Calibration and Imaging}
\label{sec:section2.2}

\begin{figure}[ht]
	\FigureFile(80mm,50mm){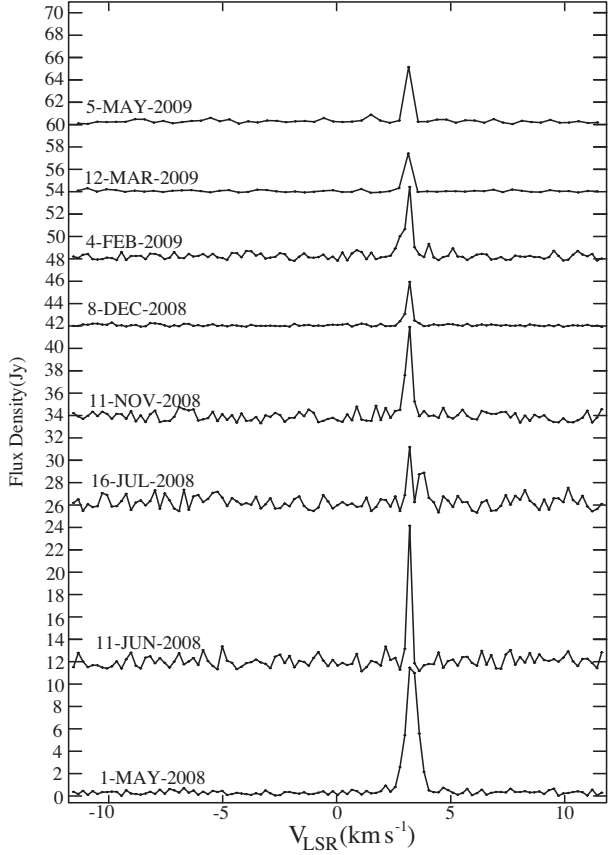}
	\caption{The time variation of the cross-power spectra of RX Boo obtained by VLBI observations at 22\,GHz with VERA. This spectra were obtained from the baseline of Mizusawa -- Iriki. We cannot show the spectrum in the 7th epoch on 2009 January 10, because Iriki station did not joined the VLBI observation.\label{fig:figure2}}
\end{figure}%

\begin{figure}[hptb]
	\FigureFile(80mm,50mm){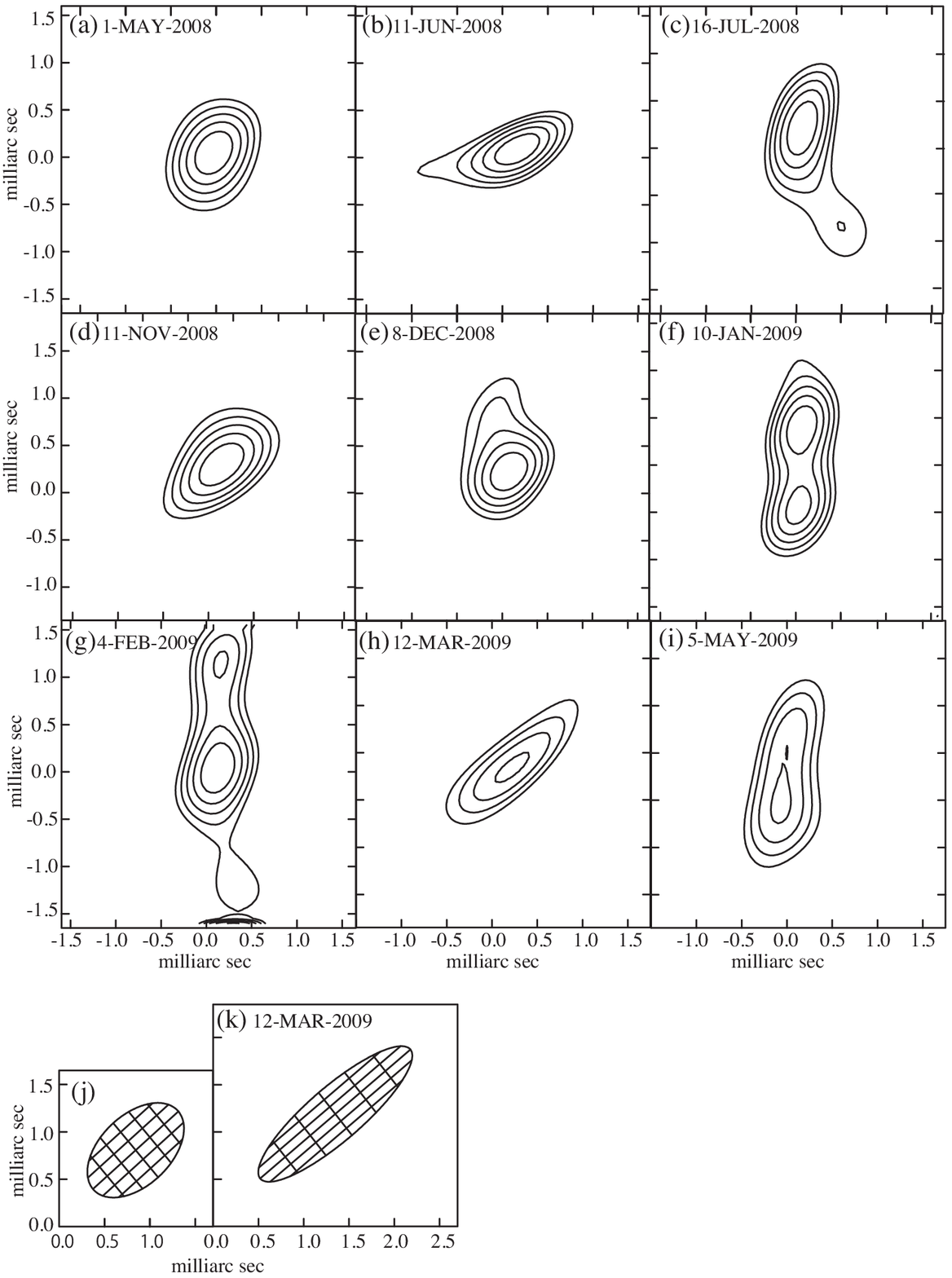}
	\caption{(a--i) Contours  of masers in  each epoch at  a channel of $V_{{\rm LSR}}=3.2 {\rm ~km\,s}^{-1}$.  The center of each map is shown in table \ref{tb:table6}. (j) Synthesized beam in the 8th epoch on 2009 February 4. The shapes of the other epochs except the 9th epoch on 2009 March 12 are almost same. (k) Synthesized beam in the 9th epoch on 2009 March 12. \label{fig:figure1}}
\end{figure}%

We analyzed VLBI data with Astronomical Imaging Package Software (AIPS) developed by National Radio Astronomical Observatory (NRAO).
We conducted the amplitude calibration using the system noise temperature recorded in each station during observations. 
In the process of fringe search of the reference source J1419+2706, we solved group delays, phases, and delay rates at intervals of 30 seconds using a task FRING with integration times of 2--3 minutes.
We transferred these solutions to the data of RX~Boo with a task TACOP and calibrated the visibilities.
We also transferred the phase and amplitude solutions derived from self-calibration of J1419+2706 and applied them to the data of RX~Boo. 

Since the a-priori delay tracking model of the correlator is not sufficiently accurate for high-precision astrometry, the recalculation of the tracking using a better model with CALC3/MSOLV \citep{Jike2005, manabe1991} were made after the correlation. 
We must correct delay- and phase-diffrences between the model of the correlator and the better model.
By multipling the delay differences and observation frequencies, the delay differences were converted to phase differences. 
These delay- and phase-differences were loaded into AIPS with a task TBIN and applied to the visibility data.
Note that the better model includes the effect of the wet atmosphere of Earth measured by the Global Positioning System (GPS) at each station \,\citep{2008PASJ...60..951H}.

To accomplish the phase referencing analysis between the target maser and continuum reference source, we calibrated phases and delay offsets derived from the differences of signal path lengths between two receivers \citep{2008PASJ...60..935H}.

As a result of these calibration, we obtained the time variation of cross-power spectra of RX~Boo as shown in  figure \ref{fig:figure2}.
From this figure, we found the LSR velocity of the peak did not change among our VLBI observations.
This component was $V_{\rm LSR}\,=\,3.2\,{\rm km\,s}^{-1}$ and had been strong before our observations, as seen in \citet{2008A&A...482..831W}.

Finally, the calibrated visibilities were Fourier transformed to make synthesized images using a task IMAGR. 
Several frequency channels at which we detected masers were imaged.
Obtained images of water masers at a LSR velosity of 3.2\,km\,s$^{-1}$ are shown in figure \ref{fig:figure1}. 
Using a task IMFIT, we fitted the image of each frequency channel to two-dimensional Gaussian models for  obtaining the positions of the maser spots. 

\section{Results}
\label{sec:section3}
\subsection{Uncertainty of each epoch}
\label{sec:section3.1}
\begin{table}[htbp]
\begin{center}
\caption{parameters of a detected maser spot in each epoch\label{tb:table6}}
\begin{tabular}{lcccccc}
\hline
epoch&$\Delta\alpha\cos\delta$[mas]$^{*}$&$\sigma_{\Delta\alpha\cos\delta}$[mas]$^{\dagger}$&$\Delta\delta$[mas]$^{*}$&$\sigma_{\Delta\delta}$[mas]$^{\dagger}$&$S$[Jy beam$^{-1}$]$^{\ddagger}$&SNR\\
\hline
2& 0.00  &0.11 & 0.00 &0.11 &6.4&25.95\\
3& $-0.52$  &0.12 &$-5.43$ &0.12 &2.3&14.32\\
4&$-2.03$  &0.15 &$-5.75$ &0.15 &4.8&11.29\\
5&$14.93$  &0.14 &$-34.18$ &0.14 &3.6&16.92\\
6&$20.02$ &0.12 &$-38.10$ &0.12 &6.2&21.33\\
7&$23.91$ &0.11 &$-41.48$ &0.11 &2.1&14.92\\
8&$26.03$ &0.13 &$-43.85$ &0.13 &2.0&14.59\\
9&$28.28$  &0.29 &$-49.64$ &0.29 &0.8&8.41\\
10&$24.56$ &0.26 &$-44.84$ &0.26 &3.4&5.38\\
\hline
\multicolumn{7}{l}{* Right Ascension and declination offsets from ($\alpha_{J2000.0}$, $\delta_{J2000.0}$)\,=\,(\timeform{14h24m11s.6206},+\timeform{25D42'12''.906}) .}\\
\multicolumn{7}{l}{$\dagger$ Position errors in right ascension and declination.}\\
\multicolumn{7}{l}{$\ddagger$ Peak fluxes of the maser spots.}\\
\end{tabular}
\end{center}
\end{table}

It is very important to determine the position error of each epoch because it affects the fitting for the annual parallax of RX~Boo.
So, we explain the position error of each epoch before least square analysis for the annual parallax.
The position error of each measurement was estimated from the root sum square of the following three error factors: 
(1) the airmass effect $\sigma_{A}$, 
(2) the errors in station positions $\sigma_{S}$, and 
(3) the quality of images $\sigma_{I}$.
After the calibration of airmass effect in the delay tracking model, there still remains an uncertainty of about 3\,cm in the zenith direction \citep{2008PASJ...60.1013N, 2007PASJ...59..889H}. 
Therefore, the error from factor (1) is estimated to be $\sigma_{A}=$80--110$\,\mu{\rm as}$ \citep{2008PASJ...60.1013N, 2007PASJ...59..889H}. 
Station positions are determined to be an accuracy of $\sim$3\,mm based on geodetic observation \citep{2008PASJ...60.1013N, 2007PASJ...59..889H}, the error from factor (2) was estimated to be $\sigma_{S} = 8\,\mu$as. 
We estimate the error from factor (3) from $\sigma_{I}=\theta_{b}/{\rm SNR}$, where $\theta_{b}$ is a root sum square of major and minor axes of the synthesized beams and SNR is a signal-to-noise ratio in the phase referenced image of the maser.
Because SNR was affected by observational condition and intensity variation of the maser, the error was estimated to be $\sigma_{I}=$50--250\,$\mu{\rm as}$. 
Thus, the error of each observation was estimated to be 110--300\,$\mu$as by taking root sum squares of these factors. 
The errors are shown in table \ref{tb:table6} and figure \ref{fig:figure3} and \ref{fig:figure4} as the error bars.
In addition to these three error factors, we can consider error raised by maser structure. 
This error factor is smaller than 110--300$\,\mu\,$as of $\sigma_{\Delta\alpha\cos\delta}$ or $\sigma_{\Delta\delta}$.
In table \ref{tb:table6}, there are positions, errors in the positions, peak Intensity $S$ and SNRs of detected maser spot in each epoch, where positions are relative offsets from the position of the maser in the 2nd epoch.

\subsection{Least square analysis for the annual parallax}
\label{sec:section3.2}

\begin{figure}[t]
	\FigureFile(80mm,50mm){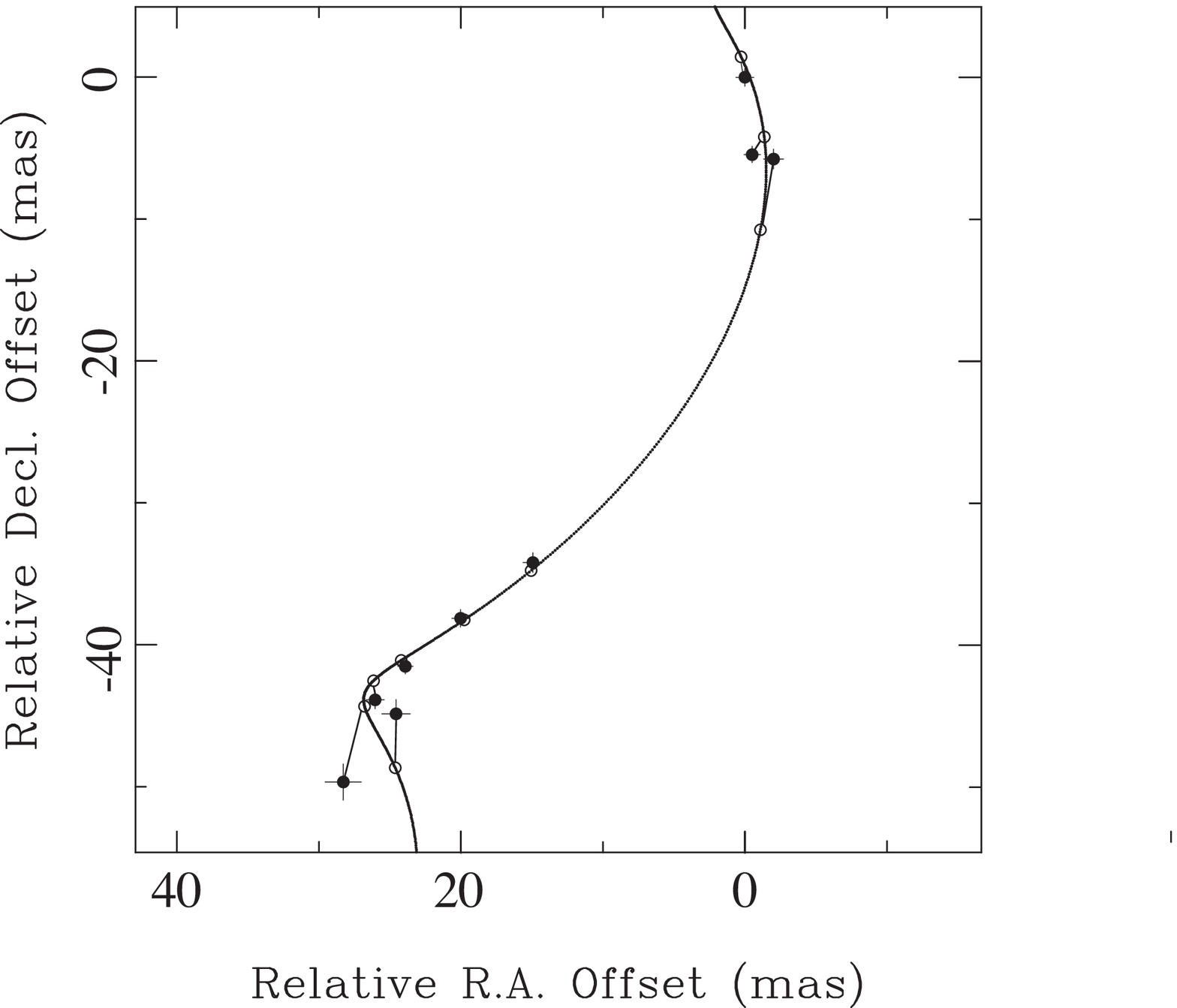}
	\caption{Relative position of the maser spot around RX~Boo with respect to the phase tracking center ($(\alpha$,\,$\delta)$=(\timeform{14h24m11s.6206},+\timeform{25D42'12''.909})). Filled circles with error bars indicate the observed positions. Open circles indicate the positions calculated from the least square fitting. 
\label{fig:figure3}}
\end{figure}%

\begin{figure}[t]
	\FigureFile(80mm,50mm){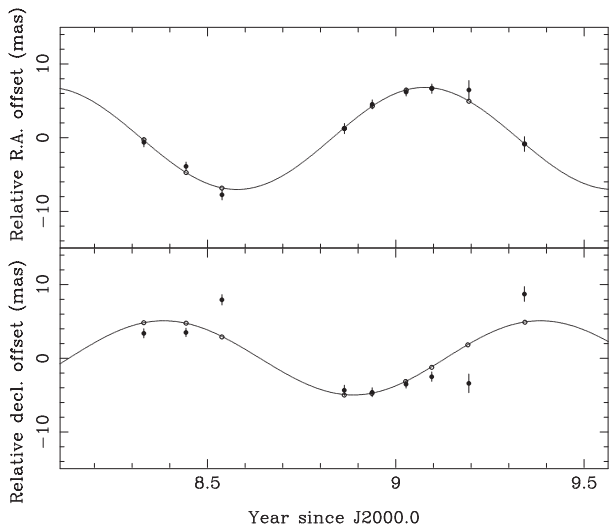}
	\caption{Time variation of the position of a maser spot around RX~Boo in right ascension (top) and declination (bottom) after subtracting the linear proper motion. The solid line is the fitting result of the annual parallax. Filled circles with error bars indicate the observed positions. Open circles indicate the positions calculated from the least square fitting.\label{fig:figure4}}
\end{figure}%

In order to obtain the annual parallax, we adopted the assumptions that the maser spot has no acceleration and the reference source is fixed on the sky.

We conducted the weighted least square fitting analysis.
Its weights are based on the position errors.
The error in each epoch is shown in table \ref{tb:table6}.
To calculate the parallax, we used maser spots which have sufficient signal-to-noise ratios on the phase-referenced images (${\rm SNR} > 5$).
Only one maser spot with $V_{\rm LSR}\,=\,3.2\,{\rm km\,s}^{-1}$ is selected with this criterion.
We performed the least square fitting by using the following equations ;
\begin{eqnarray}
\Delta\alpha\cos\delta & = & \varpi (-\sin\alpha \cos\lambda_{\odot} + \cos\epsilon\cos\alpha\sin\lambda_{\odot}) + (\mu_{\alpha} \cos\delta) t + \alpha_{0}\cos\delta\\
\Delta\delta & = & \varpi (\sin\epsilon \cos\delta \sin\lambda_{\odot} - \cos\alpha \sin\delta \cos\lambda_{\odot} - \cos\epsilon \sin\alpha \sin\delta \sin\lambda_{\odot}) + \mu_{\delta} t + \delta_{0},
\end{eqnarray}
where $(\Delta\alpha\cos\delta, \Delta\delta)$ are the displacements of the observed maser spot, ($\mu_{\alpha}\cos\delta, \mu_\delta$) are the linear motions of the maser around RX~Boo, $t$ is time, $\varpi$ is the annual parallax, $(\alpha, \delta)$ are the right ascension and declination of the source, $\lambda_{\odot}$ is ecliptic longitude of the Sun, $\epsilon$ is the obliquity of the ecliptic and $(\alpha_{0},\delta_{0})$ is the right ascension and the declination when $t = 0$.
In this fitting, we derived five parameters; $\varpi, \mu_{\alpha}\cos\delta$, $\mu_{\delta}$, $\alpha_{0}$ and $\delta_{0}$.
The total number of data is eighteen (right ascension and declination of nine epochs).

In figure \ref{fig:figure3} and \ref{fig:figure4}, we showed observed positions and results of the fitting.
In figure \ref{fig:figure3}, we present the motion of one maser spot with $V_{{\rm LSR}} = 3.2{\rm\,~km\,~s}^{-1}$ with respect to the phase tracking center. 
In figure \ref{fig:figure4}, parallactic motion in right ascension and declination are presented. 
The apparent motion clearly shows an effect of parallactic motion.
Based on the least square fitting analysis, the annual parallax was determined to be 7.31$\pm$0.50\,mas, corresponding to the distance of $136^{+10}_{-9}$ pc.
The proper motion of the spot was ($\mu_{\alpha}\cos\delta, \,\mu_{\delta})=(24.55\pm1.06\,{\rm mas\,yr}^{-1}, -49.67\pm2.38{\rm\,mas\,yr}^{-1}$).

\subsection{Comparison with previous results}
\label{sec:section3.3}
\begin{table}[htbp]
\begin{center}
\caption{our results and Hirrarcos' results\label{tb:table4}}
\begin{tabular}{lcc}
\hline
  & Our result & Hipparcos$^{*}$\\
\hline
$\varpi^{\dagger}$& 7.31$\pm$0.50\,mas  &6.42$\pm$1.00 mas \\
$\mu_{\alpha}\cos\delta^{\ddagger}$& $24.55\pm1.06\,{\rm mas\,yr}^{-1}$  &$21.74\pm0.90\,{\rm mas\,yr}^{-1}$\\
$\mu_{\delta}^{\ddagger}$&$-49.67\pm2.38{\rm\,mas\,yr}^{-1}$  &$-49.70\pm0.49{\rm\,mas\,yr}^{-1}$ \\
\hline
\multicolumn{2}{l}{* Data from \citet{1997A&A...323L..49P}}\\
\multicolumn{2}{l}{$\dagger$ Annual parallax.}\\
\multicolumn{2}{l}{$\ddagger$ proper motions.}
\end{tabular}
\end{center}
\end{table}
Here, we compare our results with previous results.
In table \ref{tb:table4}, our results and Hipparcos' results are summarized.
In comparison with Hipparcos' parallax, $\varpi_{{\rm HIP}} = 6.42 \pm 1.00 {\rm~mas}$, our parallax is consistent with the Hipparcos'  parallax.
According to Hipparcos' result, the proper motion of RX~Boo is ($\mu_{\alpha}\cos\delta, \,\mu_{\delta})=(21.74\pm0.90\,{\rm mas\,yr}^{-1},-49.70\pm0.49{\rm\,mas\,yr}^{-1}$).
In the declination component, the proper motion from our analysis is consistent with the result from Hipparcos, but in the right ascension component, the motion is $2.81\,{\rm mas\,yr}^{-1}$ larger than Hipparcos' result.
This difference of the proper motions between Hipparcos' and ours corresponds to $1.8\,{\rm km\,s}^{-1}$ at the distance of RX~Boo.
We think that this is the velocity of the maser motion with respect to the star itself.
From CO observations, the LSR velocity of the star is 1.0 ${\rm km\,s}^{-1}$ \citep{2006A&A...450..167T}.
The difference of the LSR velocities between the star and the water maser is 2.2 ${\rm km \,s^{-1}}$.
The difference of the proper motions between Hipparcos' and ours is almost as large as that of the LSR velocities between the star and the water maser.
The difference of the proper motions between Hipparcos' and ours may be caused by the motion of the maser spot with respect to the star itself.

From the directions of the proper motion and the radial velocity, we can know the positional relationship between the maser and star.
The proper motion of the maser with respect to the central star is $(\mu_{\alpha}\cos\delta, \mu_{\delta}) = (2.81{\rm~mas~yr}^{-1}, 0.03{\rm~mas~yr}^{-1})$ and directs the position angle of $\timeform{89D.4}$.
The position angle of the maser is $\timeform{89D.4}$ with respect to the central star.
We expect that this maser is located in east direction for the central star.
This is consistent with the result of VLA \citep{2008A&A...482..831W} and their detected masers which had the similar radial velocity distributed in the same direction.

\citet{2008A&A...482..831W} considered out-flow model of water maser using data observed with VLA.
They assumed the model that water masers were accelerated exponentially and expanded spherically.
Based on this assumption, they showed a relationship between expansion velocities and the three-dimensional distances from the central star to the masers;
\begin{eqnarray}
r = r_{0}-\frac{1}{k}{\rm ln}\biggl(1-\frac{v}{v_{f}}\biggr)\label{eq:equation3},
\end{eqnarray}
where $r$ is three-dimensional distance from the central star to the maser, $k = 0.065\,{\rm AU}^{-1}$, $r_{0} = 1.5\,{\rm AU}$ is the radius at which expansion velocities are zero, $v_{f} = 8.4{\rm ~km~s}^{-1}$ is the largest expansion velocity, and $v$ is the expansion velocity of masers.
By using their model, we can estimate the three-dimensional distance from central star to the maser.
On the celestial plane, the difference of proper motion between the maser and the central star is 2.8 ${\rm\,mas\,yr}^{-1}$.
This corresponds to $1.8{\rm km \,s^{-1}}$.
The difference of the radial velocities is $2.2{\rm km \,s^{-1}}$.
From these, the expansion velocity is $v = \sqrt{1.8^{2}+2.2^{2}}~=~2.8{\rm km \,s^{-1}}$.
From equation (\ref{eq:equation3}), we obtained the three-dimensional distance from central star to be 7.9 AU .
This distance is smaller than the inner radius of the detected incomplete ring structure by \citet{2008A&A...482..831W}.

\section{Discussion}
\label{sec:section4}
\subsection{Position on the PL-relation}
\label{sec:section4.1}

\begin{table}
\caption{ Period of RX~Boo\label{tb:table2}}
\begin{center}
\begin{tabular}{lccl}
\hline
\multicolumn{2}{c}{Periods (days)} &  & Reference \\ \cline{1-2}
Short Group & Long Group &  & \\
\hline
160.0        & 302.0 & & \citet{1987JBAA...97..277T}\\
179.1, 164.0 & 352.0 & & \citet{1988AN....309..323A}\\
162.3        & 304.7 & & \citet{1997ESASP.402..269M}\\
             & 340.0 & & previous version of GCVS \\
159.6        & 278.0 & & \citet{2006JAVSO..35...88S}\\
162.3  &             & & GCVS (Samus et al. (2009))\\
\hline
\end{tabular}
\end{center}
\end{table}


The parallax $7.31\pm 0.50$ mas obtained as above gives the distance modules of $-5.68$ mag. 
We apply this to assume the absolute magnitude of RX~Boo. 
For RX~Boo \citet{2007MNRAS.378.1543G} used the apparent $K$ magnitude $m_K$ of $-1.85$ mag based on the observation by Lloyd Evans and the parallax of $4.98\pm 0.64$ mas  derived from the revised Hipparcos Catalogue \citep{2005A&A...439..791V, 2007A&A...474..653V}. 
Their values gave the absolute magnitude $M_K$ of $-8.36^{+0.26}_{-0.30}$ mag. 
Our new parallax gives $M_K$ of $-7.53^{+0.14}_{-0.15}$. 
This magnitude may have also the probable error about 0.4 mag or less as indicated in \citet{2007MNRAS.378.1543G}. 

The period of variable stars classified as semi-regular (SR) is difficult to determine precisely.
In 1980s, \citet{1987JBAA...97..277T} reported the existence of two periods, 160.0 d and 302.0 d, for RX Boo.
\citet{1988AN....309..323A} obtained several periods based on their observational results.
In the General Catalogue of variable Stars (GCVS) by \citet{2009yCat....102025S} the period of 162.3~d is tabulated for RX Boo, although \citet{2007MNRAS.378.1543G} used 340~d indicated in the former version of GCVS. 
The period of 162.3 d is based on \citet{1997ESASP.402..269M}.
\citet{1997ESASP.402..269M} give the period of 304.7 d together with 162.3 d adopted in GCVS.
They have also suggested the existence of the third period of 2691.837~d without confirmation. 
A recent paper by \citet{2006JAVSO..35...88S} gives also the double-periodicity of 278.0 and 169.6~d. 
Because of the uncertainty of the period determination for the semiregular variables, 
it is difficult to judge which periods are the most essential to present the nature of the variability of RX~Boo. 
At least, we accept the fact that the stellar variability show the double-periodic nature with the shorter period of 160 - 170~d and the longer one of 280 - 350~d. 
These early studies of periods of RX~Boo are tabulated in table \ref{tb:table2}.

The period-luminosity relation is useful not only to estimate the distance of astronomical objects but also to study the stellar structure and pulsation properties. 
The above obtained $M_K$ and the periods should be compared with the PL-relation of the long-period variables in our Galaxy and the Magellanic Clouds. 
\citet{2007MNRAS.378.1543G} compared the period and $M_K$ with the Period-Luminosity sequences originally found in the Magellanic Cloud by \citet{2000PASA...17...18W} and corrected by \citet{2004MNRAS.353..705I}. 
The relation for Miras and SRa stars is labeled as sequence C, 
and that for some of SRb as sequence C$^{\prime}$. 
For the long period, RX~Boo is found on sequence C, 
and on sequence C$^{\prime}$ for the short period. 
The fact indicates that both two periods of RX~Boo look concerned with the basic properties of stellar structure not with any spontaneous occurrence. 
One of the periods corresponds to a definite mode of pulsation, and the other to another mode. 
The simultaneous enhancement of both modes may be evidence for the transient nature of RX~Boo between the sequence C and C$^{\prime}$ stars.

\subsection{Stellar radius and temperature}
The revision of the parallax will yield the derivation of the revised physical parameters of the star. 
A thorough survey of late-type giants including RX~Boo has been performed by \citet{2005A&A...442..281K}. 
They discussed the Rosseland radius for a measure of the stellar radius. 
It is defined as the radius where the optical depth calculated by using the opacities with the Rosseland mean is equal to the unity. 
The radius is very similar to the radius where the total flux corresponds to that derived from the effective temperature of stellar atmospheres. 
For RX~Boo, \citet{2005A&A...442..281K} used the angular diameter of the Rosseland radius, $18.87\pm 0.12$~mas, based on the results in \citet{1998A&A...331..619P}. 
The Rosseland radius of 278 solar radii is obtained by combining their radius and the present parallax. 
The estimated value of the radius should be increased from this when we adopt the uniform disk angular diameter of $21.0\pm 0.3$~mas derived by \citet{2002AJ....124.2821C}.

The stellar radius can be derived with another procedure that is based on the apparent magnitude, the effective temperature of stellar atmosphere, and the parallax. 
In GCVS, the maximum and minimum $V$ magntude are tabulated as 6.43 and 9.1. 
We may assume the mean $V$ magnitude of 7.8 from these values. 
\citet{2005A&A...442..281K} adopted $V$ of 7.98 in their study. 
We may assume the mean $V$ magnitude of 8.0 from the AAVSO results between 1987 and 2007 demonstrated in \citet{2008A&A...482..831W}. 
The results presented by AFOEV also show the mean $V$ magnitude of about 8.0.
After $V$ magnitude of 8.0 is chosen, the color $(V-K)$ of 9.85 is derived by using $K$ magnitude tabulated in \citet{2007MNRAS.378.1543G}.
We can obtain the bolometric correction of about $-6.5$~mag \citep{2011ApJS..193....1W}, and then we have the apparent bolometric magnitude  $m_\mathrm{BOL}$ of 1.5~mag. 
Finally, we have $M_\mathrm{BOL}$ of $-4.2$ which gives the stellar luminosity of $3630~L_\odot$ or $\log (L/L_\odot) = 3.56$. 
When we use the effective temperature of 2750~K based on \citet{2011ApJS..193....1W}, the radius of $266~R_\odot$ is obtained. 
The effective temperature of such a late M type stars is difficult to estimate from the study of stellar atmospheres. 
\citet{1999A&A...342..799A} have found that the effective temperature of semiregular variables is hotter than Mira stars. 
This result suggests that the effective temperature of RX~Boo will be higher than the mean value for the late-type giants. 
Such a higher temperature yields smaller radius for this star. 
Further study on this problem will be interesting.

At present, we may suppose the radius of RX~Boo of about $270~R_\odot$ based on the results obtained with two different procedure. 
\subsection{Stellar mass and evolutionary status}
The radius of a pulsating star can be used to estimate the stellar mass when the characteristic period, the pulsation constant, $Q$ is obtained from theoretical consideration. 
The detailed theoretical study on the radial pulsation of red ginats performed by \citet{2007MNRAS.378.1270X} gives the period ratio and the characteristic period $Q$ for various models. 
The period ratio of RX~Boo is 0.61 for the results of \citet{2006JAVSO..35...88S} and 0.53 for those of \citet{1997ESASP.402..269M}. 
The theoretical ratio of the first overtone period $P_\mathrm{1O}$ to the fundamental period $P_\mathrm{F}$ scatters for less massive and cool models.
So it looks difficult to obtain precise results but the period ratio of  0.53 is preferable for the theoretical results. 
In their theoretical results, the characteristic period for the first overtone mode $Q_\mathrm{1O}$ does not scatter even for luminous models. 
We may use $Q_\mathrm{1O}$ of about 0.039 or $\log Q_\mathrm{1O} \simeq -1.41$ when we use their algebraic expression, 
even though the luminosity of RX~Boo $\log (L/L\odot) = 3.56$ exceeds the range of their calculation unfortunately. 
From the period-density relation, 
\[
Q_\mathrm{1O} = P_\mathrm{1O}\sqrt{M/R^3},
\]
we obtain the mass of RX~Boo of 1.13\,~$M_\odot$. 
The mass estimated here will be increased when we adopt larger radius. 

The luminosity and mass of RX~Boo can be compared with the theoretical studies of AGB stars. 
A diagram titled as evolution in mass and luminosity for solar composition stars prepared by G. H. Bowen is presented in Wilson (2000). 
The diagram shows the rather schematic figure of the evolutionary path on the $\log L$-$\log M$ diagram. 
It indicates that first a star of intermediate mass evolves with the increase in the luminosity and almost constant mass, and then the mass decreases with high mass loss rate when the star reaches a critical luminosity. 
Such a critical mass-luminosity line is called as the cliff in her paper. 
The luminosity and mass obtained here for RX~Boo indicate the star on the locus of evolutionary track for the initial mass of 1~$M_\odot$ and before reaching the cliff. 
In Wilosn's diagram, the mass loss rate is also indicated, 
and the rate of $10^{-8}$~$\dot{M}$ is found for this position. 
The mass loss rate of 6~-~2~$\times 10^{-7}\ M_\odot$~ yr$^{-1}$ derived by \citet{2002A&A...391.1053O} and by \citet{2006A&A...450..167T} does not coincide with the properteis of RX~Boo presented above.

Because the derived mass depends on the adequacy of our assumption that $Q_\mathrm{1O}$ will be constant for cool and bright AGB stars, further examination on the stellar mass should be required.

\section{Summary}
\label{sec:section5}
We have measured annual parallax of the Semiregular variable RX~Bootis by tracking a maser spot at ${\mathrm V}_{{\rm LSR}}\,=\,3.2\,{\rm km\,s}^{-1}$ for one year. 
The annual parallax is 7.31$\pm$0.50\,mas, corresponding to the distance of $136^{+10}_{-9}$\,pc.
The luminosity by using the present parallax is 3630~$L_\odot$ or $\log (L/L_\odot)$ = 3.56. 
With this luminosity and latest periods, RX~Boo is found on  Sequences C and C$^{\prime}$ originally discovered on the period-luminosity diagram of the LMC. 
Based on the present results, we derived the stellar radius of about 270 $R_\odot$. 
The mass of about 1~$M_\odot$ is estimated from the radius and the theoretical pulsation properties.
This mass and the luminosity indicate that RX~Boo remains still at the slowly evolving phase based on the reults of stellar evolution consideration. 
Such an estimate matches the comparatively small mass loss rate of this star.
RX~Boo seems to be before the phase of strong mass loss.  


\end{document}